\documentclass[letterpaper, preprint, paper,11pt]{AAS}	

\usepackage{bm}
\usepackage{amsmath}
\usepackage{subfigure}
\usepackage[colorlinks=true, pdfstartview=FitV, linkcolor=black, citecolor= black, urlcolor= black]{hyperref}
\usepackage{overcite}
\usepackage{footnpag}			      	
\usepackage{amssymb}
\usepackage{algorithm,algorithmic}

\PaperNumber{18-309}

\begin{document}

\title{Data-driven framework for real-time thermospheric density estimation}

\author{Piyush M. Mehta\thanks{Assistant Professor, Department of Mechanical and Aerospace Engineering, Statler College of Engineering and Mineral Resources, West Virginia University, Morgantown, WV 26506-6106.},  
\ and Richard Linares\thanks{Charles Stark Draper Assistant Professor, Department of Aeronautics and Astronautics, Massachusetts Institute of Technology, Cambridge, MA 02139.}
}

\maketitle{}

\begin{abstract}
In this paper, we demonstrate a new data-driven framework for real-time neutral density estimation via model-data fusion in quasi-physical ionosphere-thermosphere models. The framework has two main components: (i) the development of a quasi-physical dynamic reduced order model (ROM) that uses a linear approximation of the underlying dynamics and effect of the drivers, and (ii) dynamic calibration of the ROM through estimation of the ROM coefficients that represent the model parameters. We have previously demonstrated the development of a quasi-physical ROM using simulation output from a physical model and assimilation of non-operational density estimates derived from accelerometer measurements along a single orbit. In this paper, we demonstrate the potential of the framework for use with operational measurements. We use simulated GPS-derived orbit ephemerides with 5 minute resolution as measurements. The framework is a first of its kind, simple yet robust and accurate method with high potential for providing real-time operational updates to the state of the upper atmosphere using quasi-physical models with inherent forecasting/predictive capabilities. 
\end{abstract}

\section{Introduction}\label{sec:intro}
Conjunction analysis and assessment for collision avoidance have become part of daily space operations because of the ever increasing population of resident space objects (RSOs) that constitute both operational satellites and debris. Space situational awareness (SSA) and space traffic management (STM) require active consideration in order to maintain and expand space exploration activities. In low Earth orbit, generally defined as the orbital regime spanning altitudes between 80 and 2000 km, atmospheric drag is considered the major cause of orbit prediction errors. Drag is particularly hard to model and predict because of the highly dynamic nature of the ionosphere-thermosphere (IT) system which can cause large variations in neutral mass density. While the Sun is the strongest driver of such variations, significant changes in the state of the IT system can also be caused by space weather events.

The computationally inexpensive empirical models of the thermosphere\cite{Jacchia1970,MSIS} are considered ideal for SSA/STM, however, they lack in their ability to provide accurate forecasts. The models use low-order parameterized mathematical functions that are tuned using sparse measurements and capture the variations in an average sense. The Joint Space Operation Center (JSpOC) working under the direction of the US Air Force Space Command currently uses an assimilative empirical model that makes dynamic adjustments based on recent measurements of the state of the thermosphere\cite{Storz}.

In reality, the IT is a strongly driven large-scale nonlinear physical dynamical system. The first principles based physical models appropriately use a dynamic formulation which facilitates good potential for prediction, however, realizing such a potential requires effective data assimilation or dynamic calibration methods. Incorporating observational data in the use of models has long been a critical engineering challenge. In 1960, Rudolf Kalman provided one of the most impactful solutions by developing the Kalman Filter (KF) \cite{kalman1960new}. Kalman's solution to the case with linear models and Gaussian noise has had a substantial impact on society, and for this contribution, he was awarded the 2008 Draper Prize from the National Academy of Engineering. The prize was awarded ``for the development of the optimal digital technique (Kalman filter) that is pervasively used to control a vast array of consumer, health, commercial, and defense products." However, state estimation and Uncertainty Quantification (UQ) for high-dimensional systems remains an engineering grand challenge as many approaches scale poorly with increasing dimension of the system.

Sequential state estimation methods, such as the KF and Extended Kalman Filter (EKF) \cite{optimal_ed2}, combine information from models and observations by processing observations as they become available. 
However, the KF and EKF are not easily applied to high dimensional systems since these methods do not scale well with increasing dimensions \cite{andersson2005will}. For high dimensional systems, new approaches have been developed to overcome the limitation of the KF and EKF, one such method is the ensemble Kalman filter (EnKF) \cite{evensen2003ensemble}. The EnKF uses an ensemble of simulations to compute the mean and covariance for state estimation. The EnKF has been shown to work well for various applications, \cite{evensen2003ensemble} but it still has many shortcomings \cite{andersson2005will,van2015nonlinear}. The EnKF assumes that all noise terms are Gaussian which limits its accuracy for nonlinear systems which are inherently non-Gaussian. Recent years have seen significant advances in data assimilation methods for IT models based on EnKF, however, further advances are required to reach the full potential. As a result, empirical models of the IT consistently outperform physical models in terms of accuracy because of the imperfect nature of the embedded dynamics\cite{Shim2014}.

Recent advances have been made to overcome this issue by using Particle Filters (PFs) \cite{ades2015equivalent}. PFs are nonlinear filtering approaches that solve for the pdf in a Bayesian formulation and do not make the Gaussian assumption. However, PFs do not scale well for high dimensional systems and suffer drastically from the {\em curse of dimensionality} \cite{daley1993atmospheric}. An accessible and practical engineering solution to high-dimensional systems has been to develop a Reduced-Order Model (ROM) that represents the original system using a smaller number of parameters. The Dynamic Mode Decomposition with control (DMDc) \cite{proctor2016dynamic} is a method that facilitates development of a ROM with inherent predictive/forecasting capabilities that is crucial for SSA/STM applications. Recently, the authors developed a new approach based on DMDc that exploits the Hermitian space of the problem to develop a quasi-physical ROM for thermospheric mass density from 12 years worth of physical model simulations\cite{HS-DMDc}. The authors then demonstrated data assimilation with the developed ROM using a non-operational dataset of accelerometers derived mass density\cite{DMDc-KF}. 

This paper demonstrates the simple yet robust and effective data-driven framework using simulated operational measurements for potential real-time thermospheric density estimation towards accurate density forecasts and uncertainty quantification for SSA/STM applications. The framework estimates a reduced state that represents the model parameters rather than the driver(s), which avoids degradation of the model performance in the absence of measurement data. In this paper, we demonstrate the framework using simulated GPS-derived orbit ephemerides as measurements. In addition, the quasi-physical ROM that sits at the heart of the framework can provide a 24-hour forecast in a fraction of a second on a standard desktop platform. In essence, the framework combines the best of both empirical (low cost) and physical (predictive capabilities) models. 

This paper is structured as follow: the following section provides the necessary details for developing a ROM for atmospheric mass density. Details about the methods can be found in Mehta and Linares\cite{POD_MSIS,DMDc-KF} and Mehta et al.\cite{HS-DMDc} This is followed by brief description of the dynamics model used for the orbital simulations. The following section provides details about the process of deriving the simulated orbital measurements. Next, the unscented Kalman filter technique is briefly discussed. The next section presents the results followed by conclusions. 

\section{Reduced Order Modeling}\label{s:ROM}
Even though reduced order modeling is one of the major components of the new data-driven framework demonstrated in this paper, the methods and process behind the development of reduced order models (ROM) for the IT system are well documented \cite{HS-DMDc,DMDc-KF}. Therefore, we will only provide here the basic information essential for the process of model-data fusion. 
The main idea behind reduced order modeling is to reduce the state-space dimension or number of degrees of freedom for a large-scale dynamical system. Various formulations exist for achieving this goal, each with its advantages and disadvantages. Proper Orthogonal Decomposition (POD), originally developed by Lumley\cite{Lumley}, is the most common order reduction method. One of its main drawbacks is that it does not use a dynamic formulation and requires some form of regression for model prediction\cite{POD_MSIS}. Drawing inspiration from POD, Schmid\cite{Schmid2010} overcame this limitation with Dynamic Mode Decomposition (DMD) using a dynamic formulation. Proctor et al.,\cite{proctor2016dynamic} extended the DMD formulation to systems with exogenous inputs. Building on previous work, Mehta et al.,\cite{HS-DMDc} developed that Hermitian Space-Dynamic Mode Decomposition with control (HS-DMDc) methods for batch processing of large datasets from large-scale dynamical systems. Note that POD sits at the heart of almost all new methods and developments for reduced order modeling. 

All methods rely on temporal snapshots of a systems' output to extract a reduced order representation of the underlying dynamical behavior. POD captures a significant fraction of the systems' variance/energy depending on if the decomposition is performed after/before taking away the mean. Let $\pmb \rho({\bf x},{ t})$ be the neutral density on a spatial domain defined by a uniform grid in local time, latitude, and altitude that can be decomposed into the mean ($\bar{\pmb \rho}$) and variance ($\tilde{\pmb \rho}$). The variance can be reconstructed using a finite set of characteristic spatial basis function $\Phi({\bf x})$ and the associated time-dependent coefficients ${c}({ t})$ as
\begin{equation}\label{e:POD}
\tilde{\rho}({\bf x},{ t}) = \rho({\bf x},{ t}) - \bar{\rho}({\bf x}) = \sum_{i=1}^{r} {c}_i({ t})\Phi_{i}({\bf x})
\end{equation}
where ${\bf x}$ is the spatial vector and ${t}$ is the time. The basis functions of POD modes are extracted using either an economy singular value decomposition of the snapshot matrix ${\bf X}$ (as defined below) or an economy eigen-decomposition of the correlation matrix ${\bf X}{\bf X}^T$. The snapshot matrix is computed as follows
\begin{equation}\label{e:SM_POD}
{\bf X} = \left[\begin{matrix}
| & |  &  & | \\ {\pmb \rho}_1 & {\pmb \rho}_2 & \cdots & {\pmb \rho}_{m} \\ | & |  &  & | 
\end{matrix}\right] 
\end{equation}  
where ${\bf X} \in \mathbb{R}^{n \times m}$, with $n$ being the size of the full state (the 3-dimensional grid unwrapped into a column vector) and $m$ being the number of snapshots in time.
As discussed previously, POD does not use a dynamic formulation and therefore, cannot predict ${c}({ t})$ in time.

HS-DMDc uses time-shifted snapshot matrices, in this case 12 years of simulation output from TIE-GCM covering a full solar cycle, to estimate the dynamic and input matrices of a best-fit linear dynamical system estimation:
\begin{equation}\label{e:HS-DMDc1}
{\bf X}_2 = \pmb{\mathbb{A}}{\bf X}_1 + \pmb{\mathbb{B}}{\pmb \Upsilon}
\end{equation}
where 
\begin{equation}\label{e:HS-DMDc2}
{\bf X}_1 = \left[\begin{matrix}
| & |  &  & | \\ {\pmb \rho}_1 & {\pmb \rho}_2 & \cdots & {\pmb \rho}_{m-1} \\ | & |  &  & | 
\end{matrix}\right] \quad {\bf X}_2 = \left[\begin{matrix}
| & |  &  & | \\ {\pmb \rho}_2 & {\pmb \rho}_3 & \cdots & {\pmb \rho}_{m} \\ | & |  &  & | 
\end{matrix}\right] \quad {\bf \Upsilon} = \left[\begin{matrix}
| & |  &  & | \\ {\bf u}_1 & {\bf u}_2 & \cdots & {\bf u}_{m-1} \\ | & |  &  & | 
\end{matrix}\right]
\end{equation}  
and $\textbf{u}_k$ is the input vector for time $k$. In this case, the inputs used are the solar activity proxy ($F_{10.7}$), geomagnetic proxy ($K_p$), universal time (UT) and day of the year.

In order to estimate $\pmb{\mathbb{A}}$ and $\pmb{\mathbb{B}}$, Eq. \ref{e:HS-DMDc1} is modified such that 
\begin{equation}\label{e:HS-DMDc3}
{\bf X}_2 = {\bf Z}\mathbf{\Psi}
\end{equation}
where ${\bf Z}$ and $\mathbf{\Psi}$ are the augmented operator and data matrices respectively. 
\begin{equation}\label{e:DMDc3}
{\bf Z} \triangleq \begin{bmatrix} {\bf A} & {\bf B}
\end{bmatrix} \quad \text{and} \quad \mathbf{\Psi} \triangleq \begin{bmatrix} {\bf X}_1 \\ 
\mathbf{\Upsilon}
\end{bmatrix}
\end{equation}
The estimate for ${\bf Z}$, and hence $\pmb{\mathbb{A}}$ and $\pmb{\mathbb{B}}$, is achieved with a Moore-Penrose pseudo-inverse of $\mathbf{\Psi}$ such that ${\bf Z} = {\bf X}_2\mathbf{\Psi}^{\dagger}$. 

Because the state size, $n$, can also be very large making computation and storage of the dynamic and input matrices intractable, a reduced state is used to model the evolution of the dynamical system. 
\begin{equation}\label{e:DS3}
{\bf z}_{k+1}=\pmb{\mathbb{A}}_r{\bf z}_k + \pmb{\mathbb{B}}_r{\bf u}_k + {\bf w}_k
\end{equation}
where ${\bf A}_r \in \mathbb{R}^{r\times r}$ is the reduced dynamic matrix and ${\bf B}_r \in \mathbb{R}^{r\times q}$ is the reduced input matrix in discrete time, $\textbf{z} \in \mathbb{R}^{r\times 1}$ is the reduced state, and ${\bf w}_k$ is the process noise that accounts for the unmodeled effects and the ROM truncation error. The state reduction is achieved using a similarity transform ${\bf z}_k = {\bf U}_{r}^{\dagger}{\bf x}_k = {\bf U}_{r}^{T}{\bf x}_k$, where ${\bf U}_{r}$ are the first $r$ POD modes. The steps involved in HS-DMDc are summarized below. The data assimilation process presented in this work will estimate the reduced state, $\textbf z$, that represents the coefficients of the POD modes and can be thought of as model parameters that relate the model input(s) to output(s).

\subsection{Discrete to Continuous time}\label{s:D2C}
The new framework is designed to use variants of the sequential (Kalman) filter for data assimilation that requires propagating the state(s) to the next measurement time, which most likely will not be uniformly distributed and/or with a snapshot resolution used to derive the dynamic and input matrices for the ROM. Therefore, the discrete-time dynamic and input matrices [$\pmb{\mathbb{A}}_d$, $\pmb{\mathbb{B}}_d$] need to be first converted to continuous time [$\pmb{\mathbb{A}}_c$, $\pmb{\mathbb{B}}_c$] and then back to time of next measurement, $t_k$. This can be achieved using the following relation\cite{DeCarlo}
\begin{equation}\label{e:D2C}
\begin{bmatrix} 
\pmb{\mathbb{A}}_c & \pmb{\mathbb{B}}_c\\
{\bf 0} & {\bf 0} 
\end{bmatrix} = \frac{1}{{\text T}}\log \left(\begin{bmatrix} 
\pmb{\mathbb{A}}_d & \pmb{\mathbb{B}}_d\\
{\bf 0} & {\bf I} 
\end{bmatrix}\right) 
\end{equation}
where ${\text T}$ is the sample time (snapshot resolution when converting from discrete to continuous time and the time to next measurement, $t_k$, when converting back from continuous to discrete time). This represents another major advantage of the new framework where the time-step of model evolution can be readily adjusted. 

\section{Orbital Dynamics}\label{s:OD}
In this paper, we simulate the \textit{true} orbits using 2-body dynamics with the $J_2$ and atmospheric drag perturbations. The dynamic model ${\pmb f}({\bf x},t)$ is given below:

\begin{equation}
\dot{\bf x}={\pmb f}({\bf x},t) = \begin{bmatrix}
\dot x \\ \dot y \\ \dot z \\ \dot v_x \\ \dot v_y \\ \dot v_z \\ \dot{BC} \\
\end{bmatrix}
=
\begin{bmatrix}
v_x \\ v_y \\ v_z \\
-\mu\frac{x}{r^3}-\frac{3J_2\mu R_E^2x}{2r^5}\left(1-\frac{5z^2}{r^2}\right) -  \frac{1}{2}\rho\frac{C_DA}{m} \lvert {\bf v}_{rel}\rvert v_x \\ 
-\mu\frac{y}{r^3}-\frac{3J_2\mu R_E^2y}{2r^5}\left(1-\frac{5z^2}{r^2}\right) -  \frac{1}{2}\rho\frac{C_DA}{m} \lvert {\bf v}_{rel}\rvert v_x \\
-\mu\frac{z}{r^3}-\frac{3J_2\mu R_E^2z}{2r^5}\left(3-\frac{5z^2}{r^2}\right) -  \frac{1}{2}\rho\frac{C_DA}{m} \lvert {\bf v}_{rel}\rvert v_x \\
0 \\
\end{bmatrix}
\end{equation}
where ${\bf r} = [x, y, z]$ is the inertial position, ${\bf v}_{rel} = [v_x, v_y, v_z]$ is the velocity relative to the corotating atmosphere (${\bf v}_{rel} = {\bf v} - \pmb{\omega}_E \times  {\bf r}$, where ${\bf v}$ is the inertial velocity of the spacecraft and $\pmb{\omega}_E$ is the Earth's angular velocity), $\mu$ is the Earth's gravitational parameter, $r = \sqrt{x^2+y^2+z^2}$, $\lvert {\bf v}\rvert = \sqrt{v_x^2+v_y^2+v_z^2}$, $R_E$ is the mean radius of the Earth, $J_2$ is the Earth's oblateness parameter, $\rho$ is the atmospheric mass density, and the factor $\frac{C_DA}{m}$ is the ballistic coefficient (BC).

\section{Simulated Orbits and Measurements}\label{s:SO}
Initial orbital parameters for the simulated \textit{true} orbits are sampled uniformly from the distributions provided in Table~\ref{t:Orbital_Elements_Distribution}. We restrict in this case the mean motion, in conjunction with the eccentricity, to almost circular orbits with apogee below 450 km since the current version of TIE-GCM ROM is only applicable below that altitude. We hold the BC constant and allow it to be controlled by the initial uncertainty.

\begin{table}[htb]
	\caption{Distribution of orbital parameters for \textit{true} orbits.}
	\label{t:Orbital_Elements_Distribution}
	\centering
	\begin{tabular}{r | c c}
		\hline
		\textbf{Orbital Element} &  \textbf{Minimum} & \textbf{Maximum}  \\ \hline
		\textbf{\textit{Mean Anomaly, M}}  & 0 & 2$\pi$  \\ 
		\textbf{\textit{Eccentricity, e}} & 0 & 1e-3   \\
		\textbf{\textit{RAAN, $\Omega$}} & 0 & 2$\pi$   \\
		\textbf{\textit{Argument of Perigee, $\omega$}} & 0 & 2$\pi$   \\
		\textbf{\textit{Inclination, i}} & 0 & $\pi$/2   \\
		\textbf{\textit{Mean Motion, n}} & 15.5 & 16.25   \\
		\textbf{\textit{Ballistic Coefficient, $\pmb{\frac{C_DA}{m}}$}} & 1e-3 & 1e-2   \\
		\hline
	\end{tabular}
\end{table}

\begin{figure}[h]
	\centering
	\includegraphics[width=\textwidth]{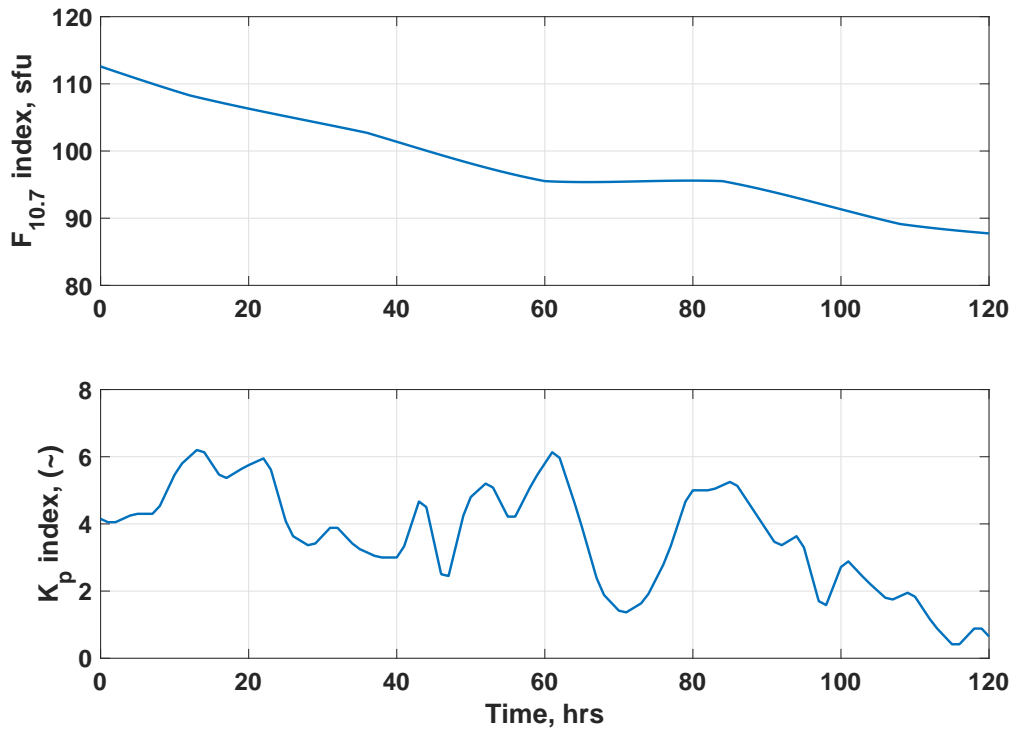}
	\caption{Time interpolated solar ($F_{10.7}$) and geomagnetic indices ($K_p$) for 5 days starting at 00:00 UT on day 191 of year 2005.}
	\label{f:Inputs}
\end{figure}

The state vectors for each simulated orbit are stacked together with the reduced state $\textbf{z}$. We randomly choose to initiate the simulation on day 191 of year 2005 at 0 UT. The TIE-GCM ROM is initialized with simulation output from TIE-GCM while the model inputs ($F_{10.7}$ and $K_p$ shown in Figure~\ref{f:Inputs}) are derived from the space weather archive on \textit{celestrak}. The initial sampled Keplerian elements are converted to inertial position and velocity and propagated for 5 days using the sampled ballistic coefficients and density from the ROM which is also simultaneously propagated as part of the full state. We propagate 10 sampled orbits and assume that each simulated orbit can be measured using a high accuracy GPS receiver on-board. We also assume that continuous GPS measurements are available with a resolution of 5 minutes. The effects of duty cycled GPS measurements will be investigated in future work. We generate the measurements by adding Gaussian noise with zero mean and a 10 m standard deviation in each dimension to the simulated \textit{true} position states. 

\section{Square Root Unscented Kalman Filter}\label{s:KF}
We use the unscented Kalman filter (UKF) for model-data fusion. The UKF was proposed by Julier and Uhlman\cite{UKF} as an extension of the very popular Kalman filter\cite{kalman1960new} for application to nonlinear systems. The UKF uses a deterministic sampling approach to avoid large errors in the true posterior mean and covariance of a Gaussian random variable (GRV) caused by first-order linearization of the nonlinear system dynamics. UKF also estimates the state distribution using a GRV, but accurately captures the true posterior mean and covariance to the 3rd order by propagating a carefully selected set of sample points through the true nonlinear system dynamics. The square root unscented Kalman filter (SQUKF) is a futher extension of the UKF for improved numerical stability. The UKF and SQUKF are very popular algorithms well documented in literature. Therefore, we will only present the SQUKF algorithm and relevant details. Description of the SQUKF presented in this paper is derived from Merwe and Wan\cite{SQUKF}.

The SQUKF uses an unscented transform (UT) to compute the statistics of a random variable that undergoes and nonlinear transformation. Let us assume a random variable $\textbf{x} \in \mathbb{R}^L$ with mean $\bar{\textbf{x}}$ and covariance $\textbf{P}_\textbf{x}$, that is propagated through a nonlinear function $\pmb f$ such that $\textbf{y} = {\pmb f}(\textbf{x})$. UT uses a set of carefully selected sample points, called sigma points, to compute the statistics of $\textbf{y}$. This is achieved by generating a matrix $\pmb{\mathcal{X}}$ of 2$L$+1 \textit{sigma vectors} $\mathcal{X}_{i}$ with corresponding weights $W_i$ and using the following relationships:
\begin{equation}\label{e:UT}
\begin{aligned}
\mathcal{X}_{0} &= \bar{\textbf{x}} \\
\mathcal{X}_{i} &= \bar{\textbf{x}} + \sqrt{(L+\lambda)\textbf{P}_\textbf{x})}_{i} \quad i=1,\dots,L\\
\mathcal{X}_{i} &= \bar{\textbf{x}} - \sqrt{(L+\lambda)\textbf{P}_\textbf{x})}_{i-L} \quad i=L+1,\dots,2L\\
W_0^{(m)} &= \lambda/(L+\lambda)\\
W_0^{(c)} &= \lambda/(L+\lambda) + (1-\alpha+\beta)\\
W_i^{(m)} &= W_i^{(c)} = 1/\{2(L+\lambda)\} \quad i=1,\dots,2L\\
\end{aligned}
\end{equation}
where $\lambda = \alpha^2(L+\kappa)-L$ us a scaling parameter. $\alpha$ determines the spread of the sigma points around $\bar{\textbf{x}}$ and $\kappa$ is a secondary scaling parameter, and $\beta$ us used to incorporate prior knowledge of the distribution of $\textbf{x}$. Based on the suggested values of the parameters and prior experience, we set the values as $\alpha = 1$, $\beta = 2$, and $\kappa = 0$.
The above computed \textit{sigma vectors} are propagated through the nonlinear function
\begin{equation}\label{e:UT1}
	\mathcal{Y} = {\pmb f}(\mathcal{X}_i) \quad i =0,\dots,2L
\end{equation}
and the mean and covariance for $\textbf{y}$ are approximated using a weighted sample means and covariance of the posterior sigma points as follows:
\begin{equation}\label{e:UT2}
\bar{\textbf{y}} \approx \sum_{i=0}^{2L} W_i^{(m)}\mathcal{Y}_i
\end{equation}
\begin{equation}\label{e:UT3}
\textbf{P}_{\textbf{y}} \approx \sum_{i=0}^{2L} W_i^{(c)} \{\mathcal{Y}_i - \bar{\textbf{y}}\} \{\mathcal{Y}_i - \bar{\textbf{y}}\}^T
\end{equation}
Both the UKF and SQUKF extend the UT to recursive estimation. The SQUKF algorithm is given below.

\begin{algorithm}
	\caption{Square Root Unscented Kalman Filter}
	
	\textbf{Initialize with:}
	
	\begin{equation}\label{e:SQUKF1}
	\hat{\textbf{x}}_0 = \mathbb{E}[\textbf{x}_0] \qquad \textbf{S}_0 = \text{chol} \{\mathbb{E}[(\textbf{x}_0-\hat{\textbf{x}}_0)(\textbf{x}_0-\hat{\textbf{x}}_0)^T]\}
	\end{equation}  
	
	\vspace{0.25cm}
	For $k \in \{1,\dots,\infty\}$,
	\vspace{0.25cm}
	
	\textbf{Sigma Point Calculation and Time Update:}
	
	\begin{equation}\label{e:SQUKF2}
	\pmb{\mathcal{X}}_k = \left[\hat{\textbf{x}}_k \quad \hat{\textbf{x}}_k \pm \sqrt{(L+\lambda)}\textbf{S}_{k}) \right]
	\end{equation}  
	
	\begin{equation}\label{e:SQUKF3}
	\pmb{\mathcal{X}}_{k+1|k} = {\pmb f}\left[\pmb{\mathcal{X}}_k, \textbf{u}_k\right]
	\end{equation}  

	\begin{equation}\label{e:SQUKF4}
	\hat{\textbf{x}}_{k+1}^- = \sum_{i=0}^{2L}W_i^{(m)}\pmb{\mathcal{X}}_{i,k+1|k}
	\end{equation}
	
	\begin{equation}\label{e:SQUKF5}
	\textbf{S}_{k+1}^- = \text{qr} \left\{\left[  \sqrt{W_1^{(c)}}  \left(\pmb{\mathcal{X}}_{1:2L,k+1|k} - \hat{\textbf{x}}_{k+1}^- \right) \quad \sqrt{\textbf{Q}}   \right]\right\} 
	\end{equation}
	
	\begin{equation}\label{e:SQUKF6}
	\textbf{S}_{k+1}^- = \text{cholupdate} \left\{ \textbf{S}_{k+1}^- , \pmb{\mathcal{X}}_{0,k+1} - 	\hat{\textbf{x}}_{k+1}^- ,  W_0^{(c)}  \right\} 
	\end{equation}
	
	\begin{equation}\label{e:SQUKF7}
	\pmb{\mathcal{Y}}_{k+1|k} = \textbf{H} \left[\pmb{\mathcal{X}}_{k+1|k}\right]
	\end{equation}

	\begin{equation}\label{e:SQUKF8}
	\hat{\textbf{y}}_{k+1}^- = \sum_{i=0}^{2L}W_i^{(m)}\mathcal{Y}_{i,k+1|k}
	\end{equation}
	
	\textbf{Measurement Update:}
	
	\begin{equation}\label{e:SQUKF9}
	\textbf{S}_{\tilde{\textbf{y}}_k} = \text{qr} \left\{\left[  \sqrt{W_1^{(c)}}  \left(\pmb{\mathcal{Y}}_{1:2L,k+1} - \hat{\textbf{y}}_{k+1} \right) \quad \sqrt{\textbf{R}}   \right]\right\} 
	\end{equation}
	
	\begin{equation}\label{e:SQUKF10}
	\textbf{S}_{\tilde{\textbf{y}}_k} = \text{cholupdate} \left\{ \textbf{S}_{\tilde{\textbf{y}}_k} , \pmb{\mathcal{Y}}_{0,k+1} - 	\hat{\textbf{y}}_{k+1} ,  W_0^{(c)}  \right\} 
	\end{equation}
	
	\begin{equation}\label{e:SQUKF11}
	\textbf{P}_{\textbf{x}_{k+1} \textbf{y}_{k+1}} = \sum_{i=0}^{2L}W_i^{(c)}   \left[\mathcal{X}_{i,k+1|k} - \hat{\textbf{x}}_{k+1}^- \right] \left[\mathcal{Y}_{i,k+1|k} - \hat{\textbf{y}}_{k+1}^- \right]^T
	\end{equation} 

	\begin{equation}\label{e:SQUKF12}
	\pmb{\mathcal{K}}_{k+1} = \left( \textbf{P}_{\textbf{x}_{k+1} \textbf{y}_{k+1}} /   \textbf{S}_{\tilde{\textbf{y}}_k}^T \right) / \textbf{S}_{\tilde{\textbf{y}}_k}
	\end{equation} 
	
	\begin{equation}\label{e:SQUKF13}
	\hat{\textbf{x}}_{k+1} = \hat{\textbf{x}}_{k+1}^- + \pmb{\mathcal{K}}_{k+1}\left(\textbf{y}_{k+1} - \hat{\textbf{y}}_{k+1}^- \right)
	\end{equation} 

	\begin{equation}\label{e:UKF14}
	\textbf{U} = \pmb{\mathcal{K}}_{k+1}\textbf{S}_{\tilde{\textbf{y}}_k}
	\end{equation} 

	\begin{equation}\label{e:UKF15}
	\textbf{S}_{k+1} = \text{cholupdate} \left\{ \textbf{S}_{k+1} , \textbf{U} ,  -1  \right\} 
	\end{equation} 
	
	\vspace{0.5cm}
	where \textbf{Q} is the process noise covariance and \textbf{R} is the measurement covariance.
\end{algorithm}

\section{Results}\label{s:R}
\begin{figure}[h]
	\centering
	\includegraphics[width=\textwidth]{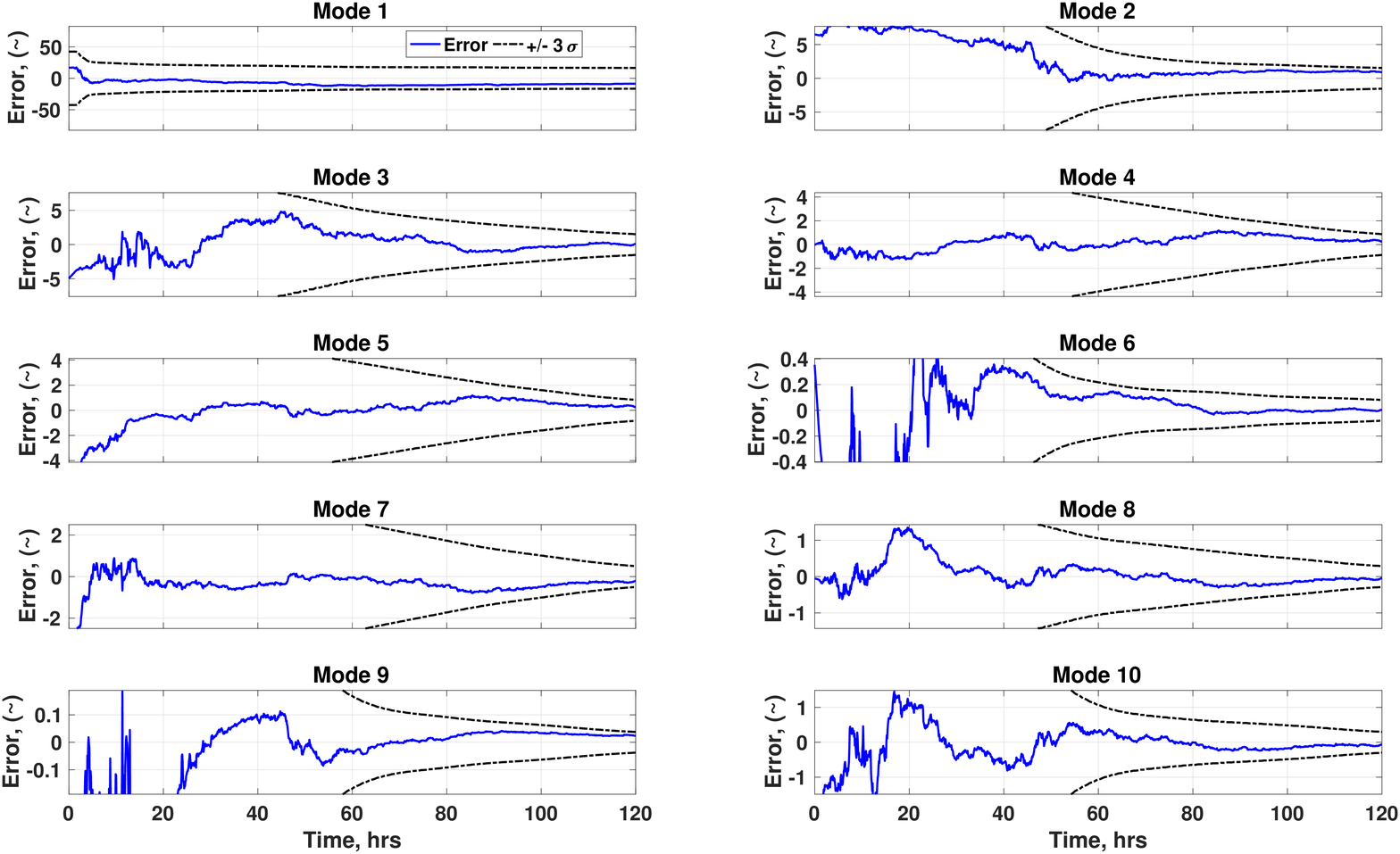}
	\caption{The error in estimated reduced order state z using 1 simulated orbit measurements since 00:00 UT on day 191 of year 2005.}
	\label{f:1Sz}
\end{figure}

The data-driven framework for real-time thermospheric density estimation is demonstrated using multiple cases, each varying by the number of simulated orbits from which measurements are available. We run cases where measurements are available from 1, 3, 5, and 10 orbital objects. For each case, the position and velocity component of the state is initialized with the \textit{true} initial position and velocity of the simulated orbits. The united for position and velocity are km and km/s, respectively. Ballistic coefficient for each assimilated orbit is perturbed by 20\% of its \textit{true} value. The ROM is initialized with the Naval Research Laboratory's MSIS (Mass Spectrometer and Incoherent Radar) model. This represents a bias/error in the state of the thermosphere with respect to the \textit{true} simulated state provided by TIE-GCM. The initial covariance ($\textbf{P}_0$) is set at the values shown below. Since the measurements are simulated with know dynamic models, we add a very small placeholder process noise ($\textbf{Q}$) as given below. Based on previous work\cite{HS-DMDc,DMDc-KF}, we use a reduced state size of $\textbf{z} \in \mathbb{R}^{10 \times 1}$ 
\begin{equation}\label{e:PQ}
\textbf{P}_0 = 
\begin{bmatrix}
P_{pos}^i \\  P_{vel}^i \\ P_{BC}^i \\ P_z^1 \\ P_z^{2:r} \\  
\end{bmatrix}
\begin{bmatrix}
1e{-4} \\ 1e{-5} \\ 2e0 \\ 2e2 \\ 2e1
\end{bmatrix} \qquad \qquad
\textbf{Q} = 
\begin{bmatrix}
Q_{pos}^i \\  Q_{vel}^i \\ Q_{BC}^i \\ Q_z^1 \\ Q_z^{2:r} \\  
\end{bmatrix}
\begin{bmatrix}
1e{-20} \\ 1e{-20} \\ 1e{-20} \\ 1e{-20} \\ 1e{-20}
\end{bmatrix}
\end{equation} 
Figure~\ref{f:1Sz} shows the errors in the estimated ROM state \textbf{z} using measurements along 1 simulated orbit. The difference between the \textit{true} (ROM) and the biased model (assumed MSIS) is clearly visible at the initial time. 
Results show that filter does relatively well in capturing the dynamics \textbf{z}$^{(3:10)}$, but falls short in terms for adjusting the absolute model bias in terms of \textbf{z}$^{(1)}$ and \textbf{z}$^{(2)}$ that represent scaling with $F_{10.7}$ after 5 days worth of measurements. The 3$\sigma$ uncertainty bounds remain rather large even after 5 days worth of measurements. The results suggest an ambiguity problem when using measurements along a single orbit. In other words, multiple combinations of density and BC may be able to provide the same solution along a single orbit.

\begin{figure}[h]
	\centering
	\includegraphics[width=\textwidth]{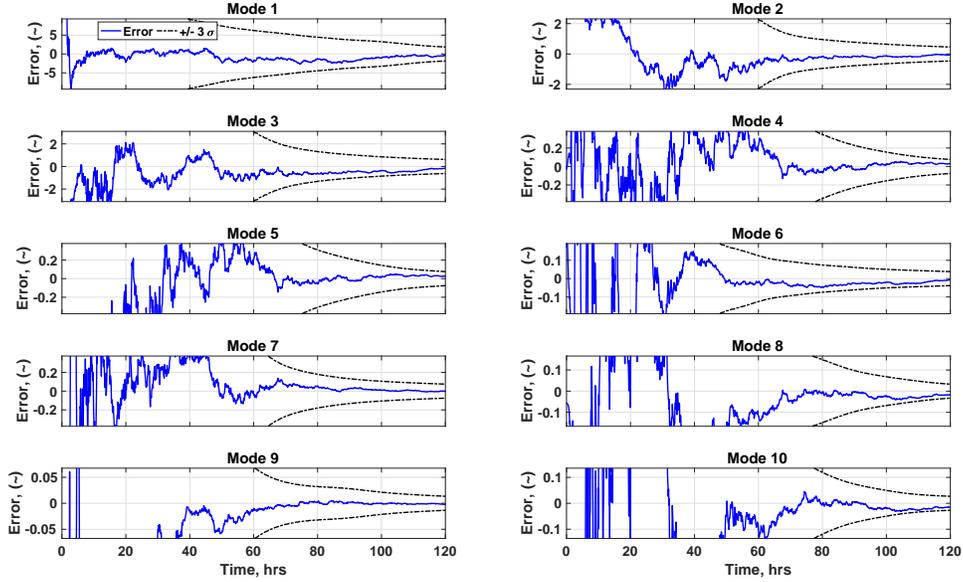}
	\caption{The error in estimated reduced order state z using 3 simulated orbit measurements since 00:00 UT on day 191 of year 2005.}
	\label{f:3Sz}
\end{figure}

Figure~\ref{f:3Sz} shows the errors in the estimated ROM state \textbf{z} using measurements along 3 simulated orbits. In contrast to the case with measurements along 1 simulated orbit, the 3 orbit case shows better agreement between the \textit{true} and estimated state for $\textbf{z}^{(1)}$ and $\textbf{z}^{(2)}$ with significantly smaller 3$\sigma$ bounds for $\textbf{z}^{(1)}$. This is because measurements along 3 randomly significantly reduces the possibility of an ambiguous solutions. The errors still suggest a small bias in $\textbf{z}^{(1)}$ after 5 days, however, the declining tends also suggest a possibility of convergence with assimilation of more data. 

\begin{figure}[h]
	\centering
	\includegraphics[width=\textwidth]{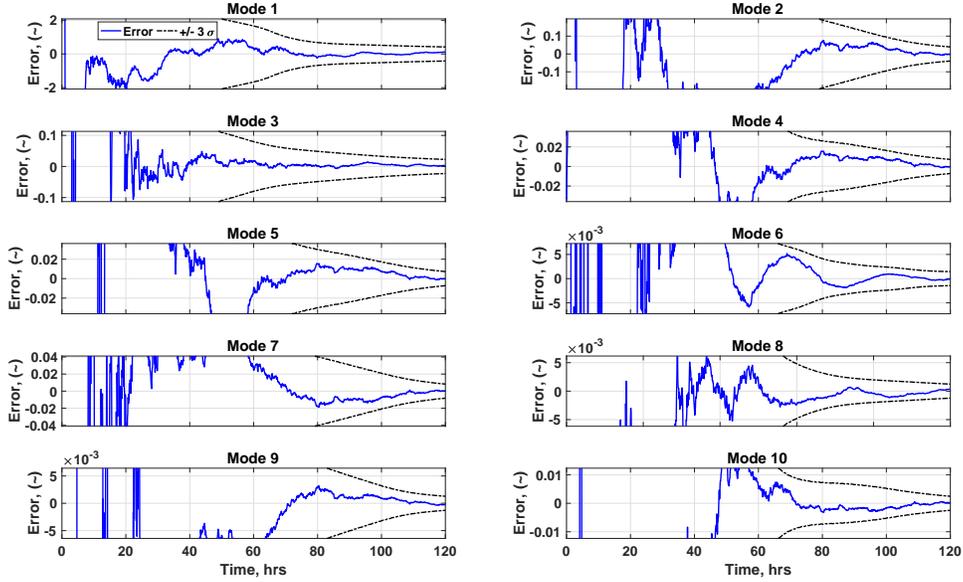}
	\caption{The error in estimated reduced order state z using 5 simulated orbit measurements since 00:00 UT on day 191 of year 2005.}
	\label{f:5Sz}
\end{figure}

\begin{figure}[h]
	\centering
	\includegraphics[width=\textwidth]{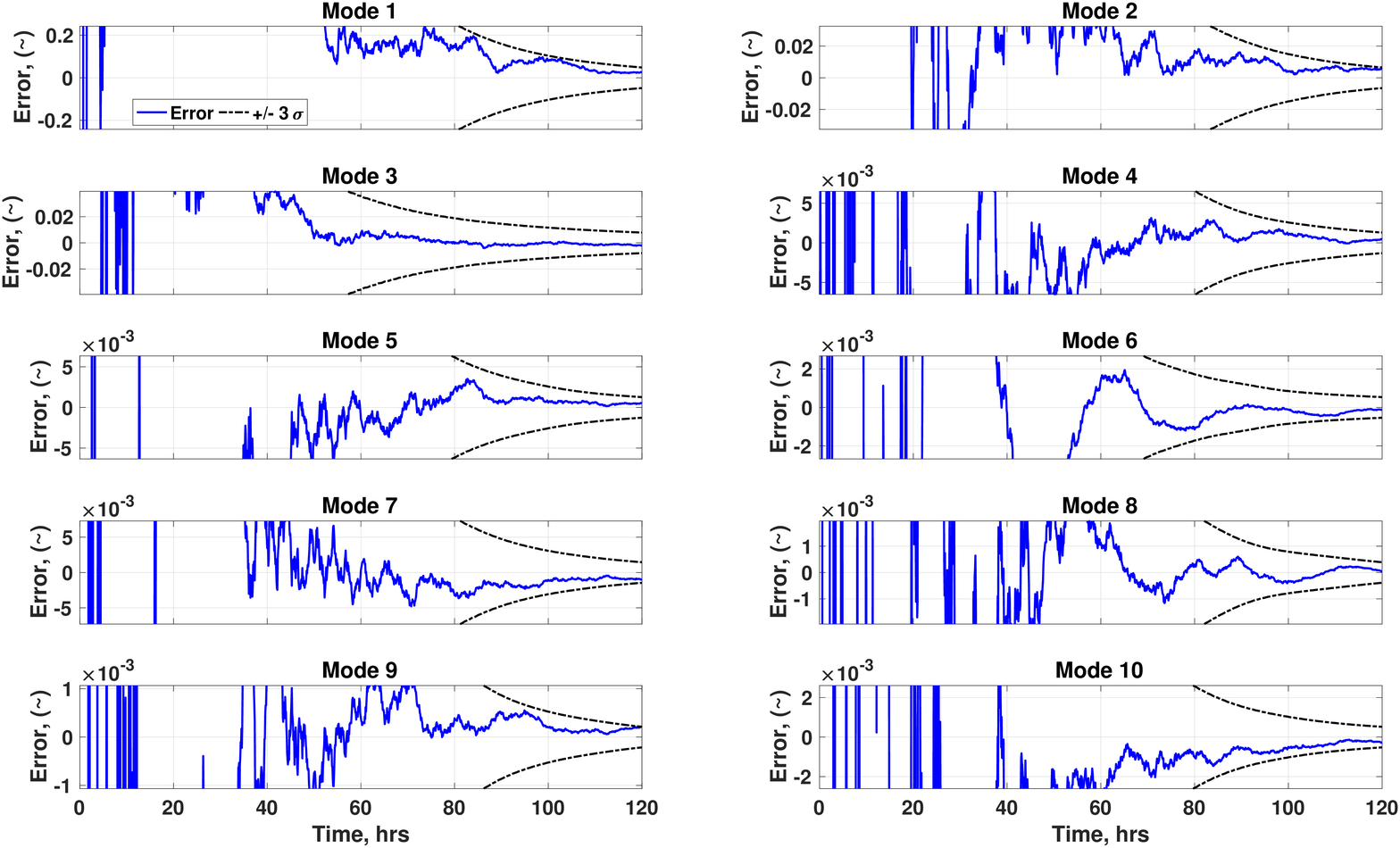}
	\caption{The error in estimated reduced order state z using 10 simulated orbit measurements since 00:00 UT on day 191 of year 2005.}
	\label{f:10Sz}
\end{figure}

\begin{figure}[h]
	\centering
	\includegraphics[width=\textwidth]{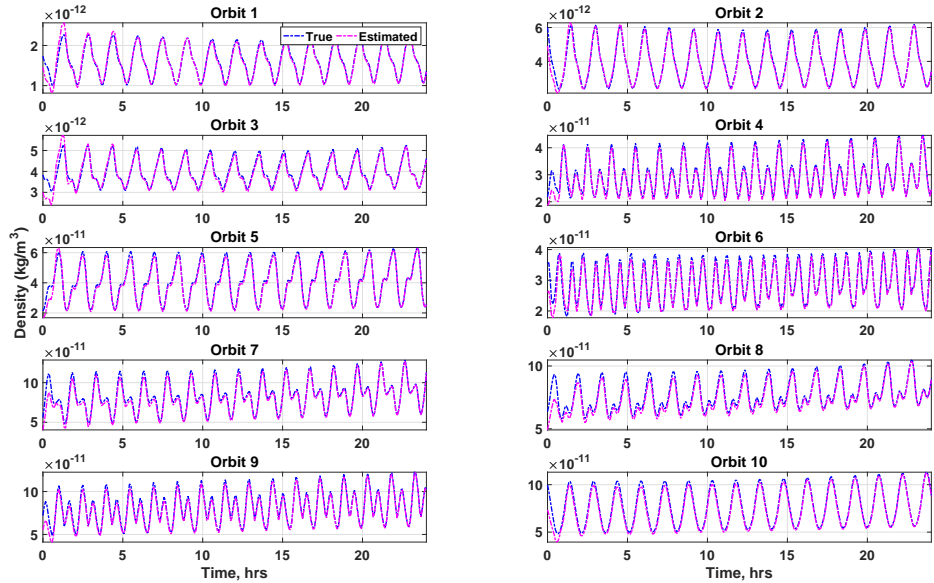}
	\caption{The estimated density along true orbit(s) using 10 simulated orbit measurements since 00:00 UT on day 191 of year 2005. blue: the \textit{true} density used in the generation of the simulated measurements. magenta: density from UKF estimated state.}
	\label{f:10SD}
\end{figure}

\begin{figure}[h]
	\centering
	\includegraphics[width=\textwidth]{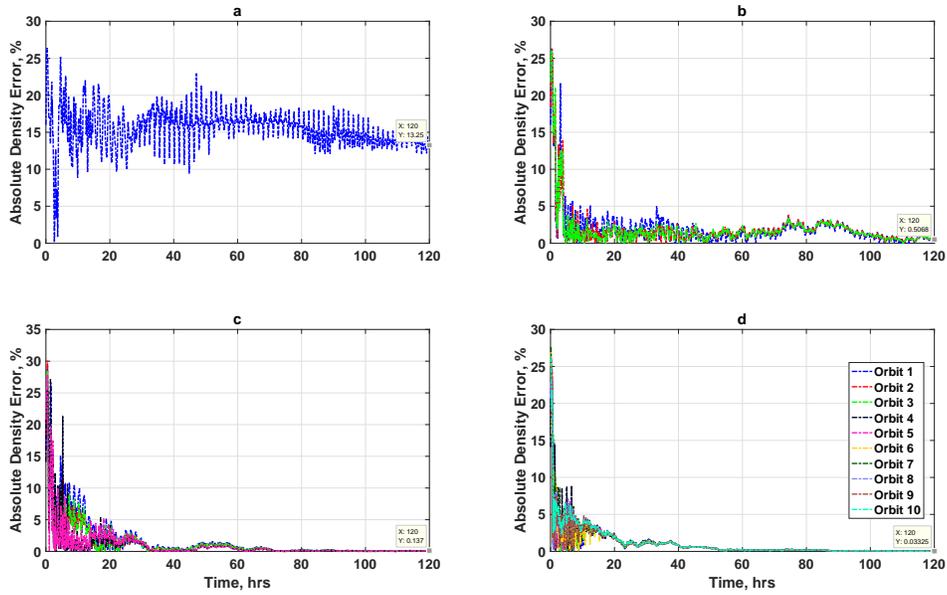}
	\caption{Error in density along the simulated orbits when assimilating measurements along (a) 1, (b) 3, (c) 5, and (d) 10 orbit(s).}
	\label{f:Den_Er}
\end{figure}

\begin{figure}[h]
	\centering
	\includegraphics[width=\textwidth]{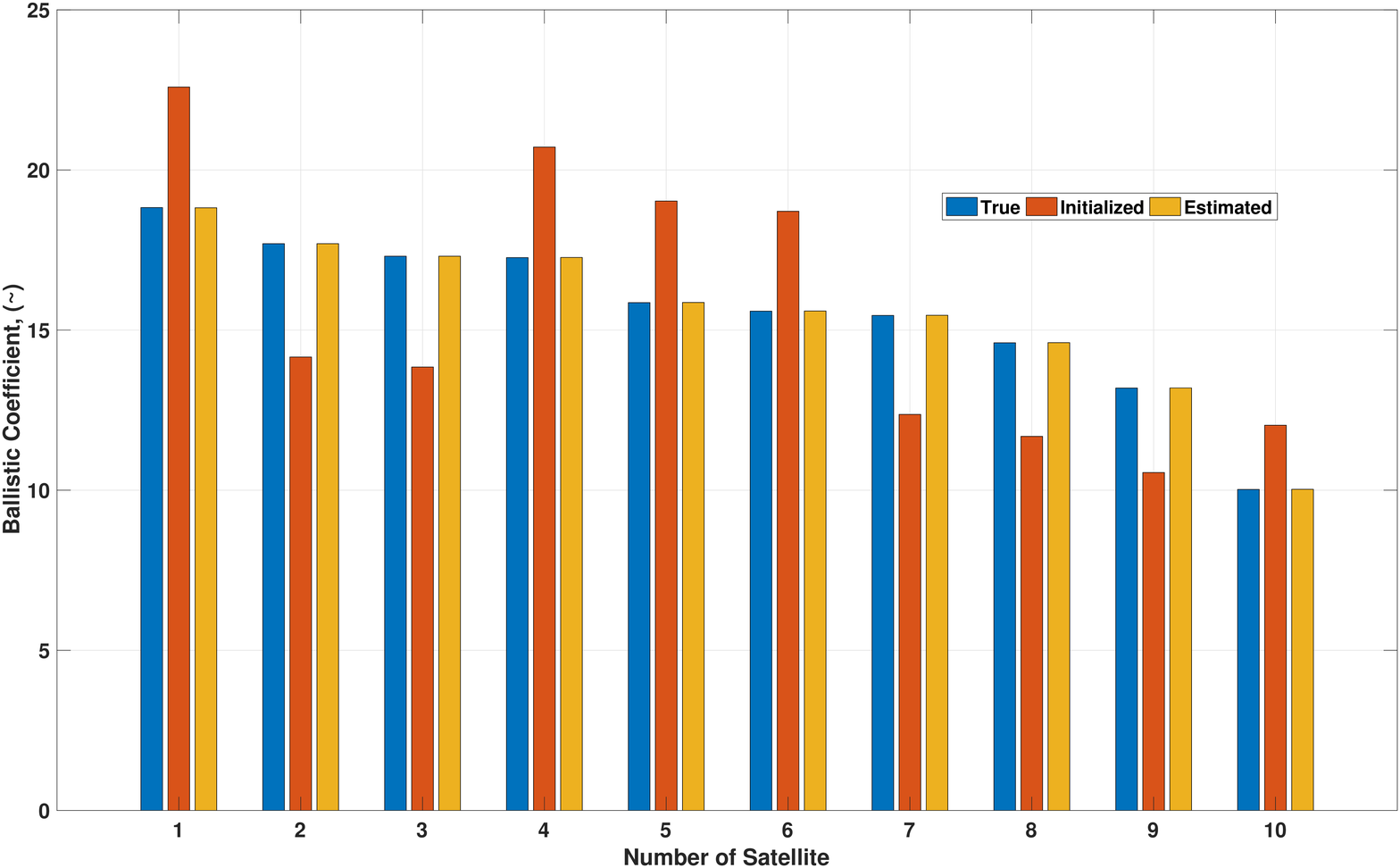}
	\caption{Comparison of the \textit{true} (blue), initial (red), and estimated (yellow) ballistic coefficients for the 10 simulated orbit.}
	\label{f:10SB}
\end{figure}

Figure \ref{f:5Sz} and \ref{f:10Sz} shows the performance of the framework using measurements from 5 and 10 simulated orbits, respectively. The 3$\sigma$ continue to shrink with increasing number of measurement orbits. Results for both cases suggest good model-data convergence for the full reduced state. Since the results suggest convergence using 10 simulated orbits, we plot the \textit{true} and estimated densities along the 10 orbits for the first 24 hours in Figure \ref{f:10SD}. The difference in \textit{true} and estimated densities for all the cases is shown in figure \ref{f:Den_Er}. The difference in densities at initial time suggest altitude dependent bias between the models (ROM initialized with TIE-GCM and MSIS). The observed differences for the 3 orbit case are significantly smaller than the 1 orbit case, while the difference approaches zero after 5 days as the number of simulated orbits increases. The comparison suggests convergence of the estimated and \textit{true} state with 10 simulated orbits. Figure \ref{f:10SB} shows the comparison of the \textit{true}, initial, and estimated ballistic coefficients for the 10 orbits. The estimated ballistic coefficients approach the \textit{true} values even with perturbations in the initial values of 20\%. The framework shows the promise of self-consistently correcting the state of the thermosphere bringing it closer to the \textit{true} state.

\begin{figure}[h]
	\centering
	\includegraphics[width=\textwidth]{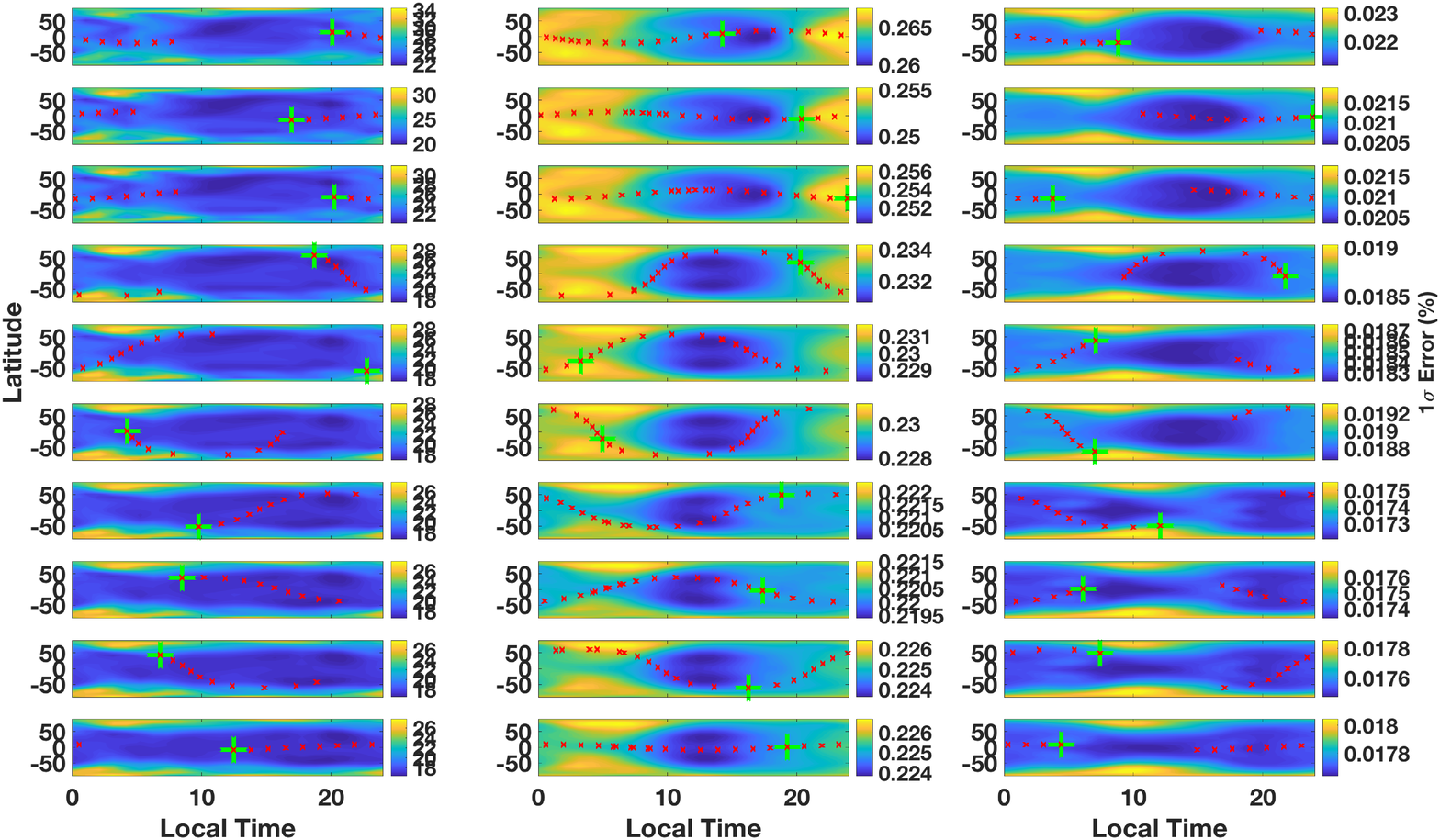}
	\caption{Projected uncertainty in density at instantaneous altitudes of the 10 simulated orbits at initial time (left column), two and half days through the five day period (middle column), and at the end of the 5 day period (right column). The red markers show the orbit about the current satellite location shown with the green marker.}
	\label{f:ER_CO}
\end{figure}

Table \ref{t:Orbital_Elements} gives the initial orbital parameters for the 10 simulated \textit{true} orbits. The 1, 3, and 5 orbit cases uses the first, first three, and first five orbits in table \ref{t:Orbital_Elements}, respectively. Figure \ref{f:ER_CO} shows the estimated uncertainty in density projected onto the latitude-local time plane at the instantaneous altitudes of the 10 simulated orbits at initial time (left column), two and half days through the five day period (middle column), and at the end of the 5 day period (right column). Results show that assimilating data along only 10 orbits provides a global reduction of uncertainty in density. The uncertainty starts high with a latitude-local time structure but reduces to almost a constant low level post data assimilation. Note the very tight scales of the 1$\sigma$ errors in middle and right columns.

\begin{table}[htb]
	\caption{Orbital parameters for the 10 simulated \textit{true} orbits.}
	\label{t:Orbital_Elements}
	\centering
	\begin{tabular}{r | c c c c c c c}
		\hline
		\textbf{Parameter} $\rightarrow$ &  \textbf{\textit{M}} & \textbf{\textit{e}} & \textbf{\textit{$\Omega$}} & \textbf{\textit{$\omega$}} & \textbf{\textit{i}} & \textbf{\textit{n}} & \textbf{\textit{BC}}\\ \hline
		\textbf{\textit{Orbit 1}}  & 1.579 & 1.076e-4 & 1.674 & 0.720 & 0.350 & 1.130e-3 & 18.825 \\ 
		\textbf{\textit{Orbit 2}} & 5.660 & 7.729e-4 & 4.612 & 5.475 & 0.216 & 1.142e-3 & 18.829 \\
		\textbf{\textit{Orbit 3}} & 3.075 & 7.044e-4 & 0.271 & 0.712 & 0.242 & 1.142e-3 & 17.307 \\
		\textbf{\textit{Orbit 4}} & 5.593 & 2.822e-4 & 1.082 & 2.723 & 1.236 & 1.162e-3 & 17.263 \\
		\textbf{\textit{Orbit 5}} & 2.601 & 9.462e-4 & 6.047 & 2.237 & 1.006 & 1.165e-3 & 15.857 \\
		\textbf{\textit{Orbit 6}} & 4.379 & 8.129e-4 & 3.006 & 5.007 & 1.287 & 1.166e-3 & 15.592 \\
		\textbf{\textit{Orbit 7}} & 4.776 & 9.737e-4 & 2.495 & 0.168 & 0.932 & 1.172e-3 & 15.592 \\
		\textbf{\textit{Orbit 8}} & 4.094 & 6.304e-4 & 6.073 & 3.445 & 0.672 & 1.173e-3 & 14.603 \\
		\textbf{\textit{Orbit 9}} & 0.630 & 9.947e-4 & 4.204 & 1.618 & 1.064 & 1.173e-3 & 13.190 \\
		\textbf{\textit{Orbit 10}} & 1.372 & 9.005e-4 & 3.408 & 3.506 & 0.151 & 1.174e-3 & 10.024 \\
		\hline
	\end{tabular}
\end{table}

\begin{figure}[h]
	\centering
	\includegraphics[width=\textwidth]{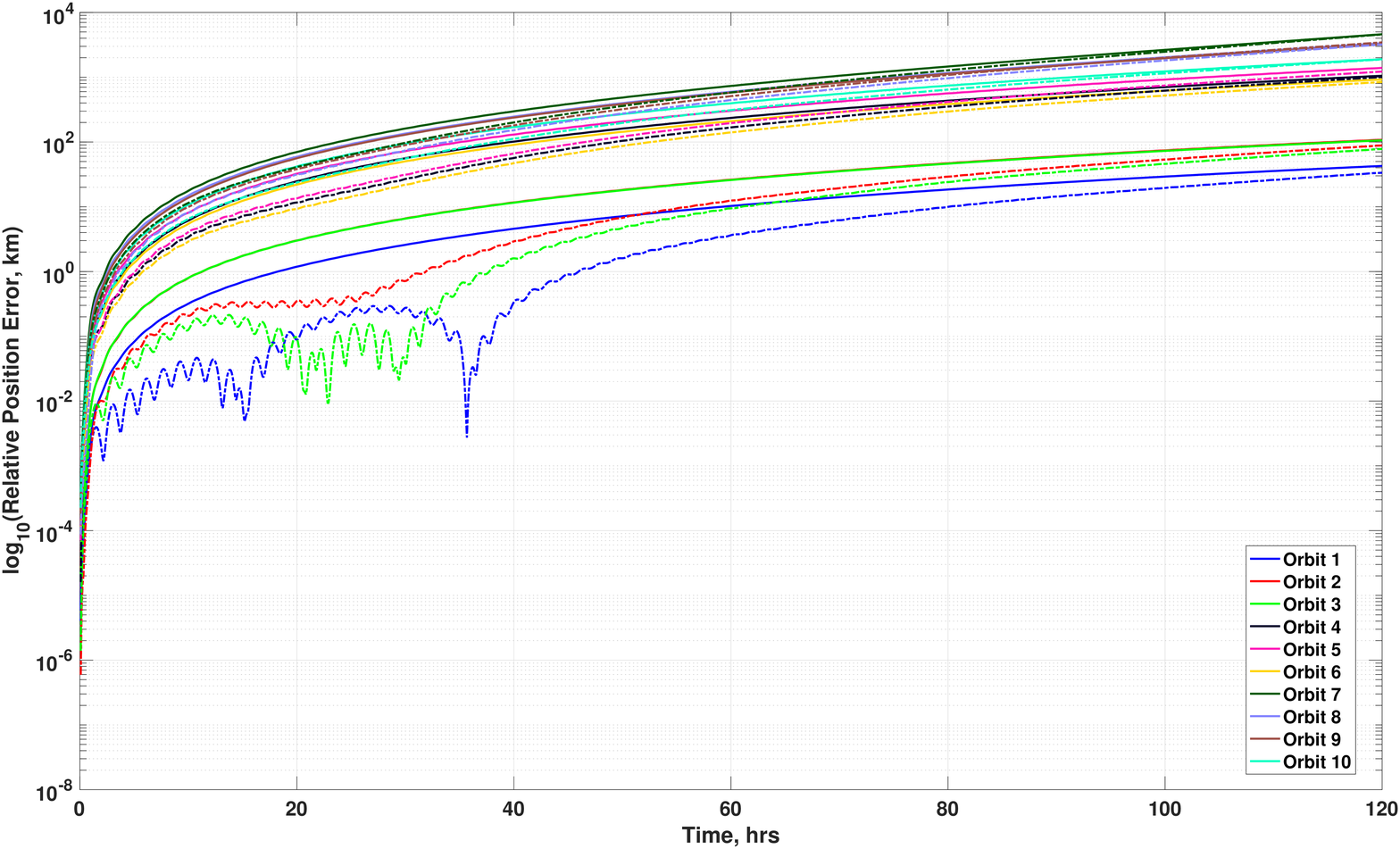}
	\caption{Orbit prediction errors for the 10 simulated orbits since 00:00 UT on day 191 of year 2005 when using density from (solid) ROM initialized with MSIS and (dotted) MSIS. Densities for the \textit{true} simulated orbits are computed with the ROM initialized with TIE-GCM.}
	\label{f:RPE}
\end{figure}

Figure \ref{f:RPE} shows the orbit prediction errors for the 10 simulated orbits when using MSIS for density, highlighting the need for an efficient framework for data assimilation of (quasi-)physical thermosphere models. The solid lines show the errors when using densities from ROM initialized with MSIS, whereas the dotted lines show the errors when using densities from MSIS. The minimum error after 72 hours is close to 6 km whereas the largest error after 72 hours stand close to 1000 km. The minimum and maximum errorr increase to more than 30 km and 3000 km after 120 hours, respectively. The large errors correspond to orbits with large BC and/or low altitudes for which drag acceleration is significant.

%

\section{Conclusions}
\label{s:con}

Atmospheric drag remains the large source of uncertainty in orbit prediction for collision avoidance and re-entry prediction for objects that traverse low Earth orbit. The current state-of-practice uses an assimilative empirical model that makes dynamic adjustments based on recent measurements of the state of the thermosphere. The empirical formulation in fast and ideal for space situational awareness/space traffic management application but inherently lacks predictive/forecasting capabilities. Physical models on the other hand use a dynamics formulation with good potential for predictive capabilities, however, they can be computationally expensive and more importantly require development of effective data assimilation methods to reach their full potential.

Recently, the authors have proposed and developed a new framework based model order reduction towards a reduced order representation of physical models that is fast to evaluate and possesses inherent predictive capabilities. In addition, the framework also significantly simplifies the process of data assimilation or model-data fusion by reducing the dimension of the state to a handful of parameters. The authors have already presented the development of a reduced order model for thermospheric mass density based on 12 years of simulation output from a physical model. They have also previously demonstrated the ability of the framework for effective and efficient data assimilation using accelerometer derived non-operational datasets.

In this work, we demonstrate the potential of the framework for dynamic calibration in real-time using simulated operational data. We use orbits simulated with 2-body, $J_2$, and drag as measurements towards estimation of the \textit{true} (simulated) state of the thermosphere. Results shows that the framework has good potential for effective and dynamic calibration of the upper atmosphere in real-time using measurements along only 10 spatially distributed orbits. 

Continuous availability of GPS-derived orbit measurements at a 5 minute resolution is assumed for this paper and the effect of GPS duty cycle on the performance of the framework is left for future work. In addition, future work will also investigate and evaluate the performance of framework using other operational datasets such as radar and/or TLEs.

\section{Acknowledgment}
The authors wish to acknowledge support of this work by the Air Force's Office of Scientic Research under Contract Number FA9550-18-1-0149 issued by Erik Blasch. MSIS data used in this work are derived from the model downloaded at https://www.brodo.de/space/nrlmsise/. The authors wish to thank Eric Sutton of the Air Force Research Laboratory for providing the TIE-GCM simulation output data used in this work.

\bibliographystyle{AAS_publication}   
\bibliography{references}   

\end{document}